\documentstyle[psfig,12pt,qqa4lart]{article}
\input tcilatex
\begin{document}


\QQQ{language}{
American English}

\author{V. R. Chitnis and P. N. Bhat \\ 
{\it Tata Institute of Fundamental Research,} \\{\it Homi Bhabha Road, Mumbai 400 005, India.}}

\title{Simulation Studies on Arrival Time Distributions of \v Cerenkov Photons
in Extensive Air Showers}
\maketitle

\begin{abstract}
Atmospheric \v Cerenkov technique is an established methodology to study
TeV energy gamma rays.  Here we carry out systematic monte carlo
simulation studies of the timing information of \v Cerenkov photons.
Extensive studies have already been carried out in this regard. Most of
these are carried out at higher energies with the aim of studying
the elemental composition of cosmic rays.  However not
much attention is paid to the species dependent signatures at TeV
energies.  In this work, functional fits have been carried out to
the spherical \v Cerenkov shower fronts and the radii of curvature have
been found to be equal to the height of shower maximum irrespective of the
species or the observation level. Functional fits have also been carried
out to describe the pulse shapes at various core distances in terms of
well known probability density distribution functions (PDF). Two types of PDF's
have been tried viz.  $\Gamma -$function and lognormal function. The
variation of the pulse shape parameters as a function of primary energy,
observation height and incident angles have been studied. The possibility
of deriving the pulse shape parameters like the rise \& decay times, full
width at half maximum from the easily measurable quantities like the mean
and RMS variation of photon arrival times offers a very important new
technique which can be easily applied in an observation.

\end{abstract}
\vskip 0.25in

\section{Introduction}

Atmospheric \v Cerenokov Technique (ACT), is by now a well established and
a unique method for the astronomical investigation of Very High Energy
(VHE, also referred to as TeV) $\gamma -$ rays. It is mainly based on the
effective detection and study of the \v Cerenkov light emitted by the
secondary particles produced in the extensive air showers initiated by the
primary $\gamma -$ ray [1, 2, 3].

Extensive studies have already been carried out on the temporal structure
of \v Cerenkov photons arriving at the observation level using detailed
simulation techniques.  Most of these studies are carried out at higher
energies with the aim of studying the elemental composition of cosmic rays
at these energies. The potential of the measurements of this radiation for
giving an insight into the longitudinal cascade development of large
showers has already been realised. These measurements have been exploited
in estimating the mean atomic mass number of the primary particles with
energies in excess of $10^{17} eV$ [4].

It has been noticed, however, that there are not many results available on
the \v Cerenkov photon arrival time studies at lower primary energies {\it e.g.}
a few hundred GeV. In the present work, we carry out a systematic
study of the temporal and spatial profile of \v Cerenkov light from lower
energy primaries both from pure electromagnetic cascades as well as
hadronic cascades. Variations of experimentally measurable temporal
parameters are studied both for electromagnetic as well as hadronic
showers of varying primary energies and incident angles. Core distance,  
altitude and incident angle dependences are also systematically studied with 
the primary aim of identifying species dependent differences in the temporal 
structure of the \v Cerenkov light.

\section{Simulations}

\v Cerenkov light emission in the earth's atmosphere by the secondaries of
the air shower generated by cosmic ray primaries or $\gamma -$ rays have
been simulated using a package called CORSIKA [5]
version 560. This program simulates interactions of nuclei, hadrons,
muons, electrons and photons as well as decays of unstable secondaries in
the atmosphere. It uses EGS4 code [6] for the electromagnetic
component of the air shower simulation and dual parton model for the
simulation of hadronic interactions. The \v Cerenkov radiation produced
within the specified band width (300-650 nm) by the charged secondaries is
propagated to the ground.  The US standard atmosphere parameterized by
Linsley [7] has been used in our simulation. The position,
angle, time (with respect to the first interaction) and production height
of each photon hitting the detector on the observation level are recorded.

       In the present simulation studies we have used Pachmarhi
(longitude: 78$^{\circ}$ 26$^{\prime}$ E, latitude: 22$^{\circ}$
28$^{\prime} N$ and altitude: 1075 $m$) as the observation level where an
array of \v Cerenkov detectors each of area $4.35~m^2$. is deployed in the
form of a rectangular array. For simulations we have used much larger
array consisting of 17 detectors in the E-W direction with a separation of
25 m and 21 detectors in the N-S direction with a separation of 20 m.
This configuration similar to the Pachmarhi Array of \v Cerenkov
Telescopes (PACT) is chosen so that one can study the core
distance dependence of various observable parameters [8]. Monoenergetic
primaries consisting of $\gamma -$ rays, protons and iron nuclei incident
vertically on the top of the atmosphere with their cores at the centre of
the array have been simulated in the present studies.

	An option of variable bunch size of the \v Cerenkov photons is
available in the package which serves to reduce the requirement of
hardware resources. However since we are interested in the fluctuations of
each of the estimated observables we have tracked single photons for each
primary at all energies. Multiple scattering length for $e^+$ and
$e^-$ is decided by the parameter STEPFC in the EGS code which has
been set to 0.1 in the present studies [9]. However
wavelength dependent absorption of \v Cerenkov photons in the atmosphere
is not taken into account. All the \v Cerenkov photons arriving at a detector
irrespective of their incident angles are accepted in the present studies.

\section{Shape of shower front}

We have studied the arrival time distribution of \v Cerenkov photons as a
function of distance from the core of the shower for both $\gamma -$ rays
and protons. We approximate the \v Cerenkov light detected at the
observation level to a spherical front moving with the speed of light {\it
c}, originating from a fixed point on the shower axis [10].
As we will see later this point happens to be at the point of
shower maximum. For vertical showers the relative time delay {\it t(r)} at
a core distance {\it r} is approximated by :

\begin{equation}
t(r) = {\sqrt {(R^2+r^2)} \over {c}} - {{R} \over {c}} 
\end{equation}

where {\it R} is the radius of curvature of the spherical front
[11]. This equation basically approximates the origin
of \v Cerenkov light to a single point on the shower axis at a height {\it
R} above the observation level with respect to which time is measured.

Around 50, 20 \& 9 showers were generated for $\gamma -$ rays of energy 100 GeV,
500 GeV and 1 TeV respectively and also for protons of energies 250 GeV, 1 TeV 
and 2 TeV
incident vertically at the top of the atmosphere. Similarly, 34 \& 10 showers
were generated for Fe nuclei of energy 5 and 10 TeV respectively. The shower 
core is
always chosen to be the centre of the array.  Energies of $\gamma -$ rays
and protons are chosen so that they have comparable \v Cerenkov yields.
For each shower, arrival time is measured with respect to the first photon
hitting the array. Variation of the arrival time with respect to the core
distance, averaged over specified number of showers, is shown in figure 1.  
It can be seen
that the wavefront has a clear spherical shape compared to the shower
front of EAS particles. Conventionally, the EAS shower fronts are fitted to
a cone with its axis coinciding with the shower core [12,13]
while it would give a poor
approximation to the \v Cerenkov front. In addition, the thickness of the
\v Cerenkov disk is much less compared that of the EAS particle disk,
especially at large core distances. As a result, the \v Cerenkov disk with
its low timing jitter and a well defined shape is better suited for the
measurement of  arrival angle of the EAS.

 The delay profiles shown in figure 1 are fitted to a spherical wavefront
and the fitted radii of these wavefronts are listed in Table 1. These
roughly correspond to the position of the shower maximum, which is
expected since a large fraction of the \v Cerenkov emission comes from the
regions in the vicinity of shower maximum [14]. It can be observed from
table 1 that the fitted radii for a given species decrease with increasing
primary energy. This is mainly due to the well known fact that higher
energy primaries propagate farther from the first interaction point before
reaching the shower maximum. This is not strictly true in the case of hadronic 
primaries because of large shower to shower fluctuations. Also a good 
anti-correlation is seen between
the total number of detected \v Cerenkov photons from  monoenergetic
primaries and the fitted radii.  
This is perhaps understandable because a larger photon
number in a given shower is also due to an increase in the primary energy.  
Such a correlation is weak in the case of hadronic primaries. 
The intrinsic correlations are smeared due to large
fluctuations in the case of hadronic primaries. These results are
consistent with those obtained in an earlier work [14].

\begin{figure}
\centerline{\psfig{file=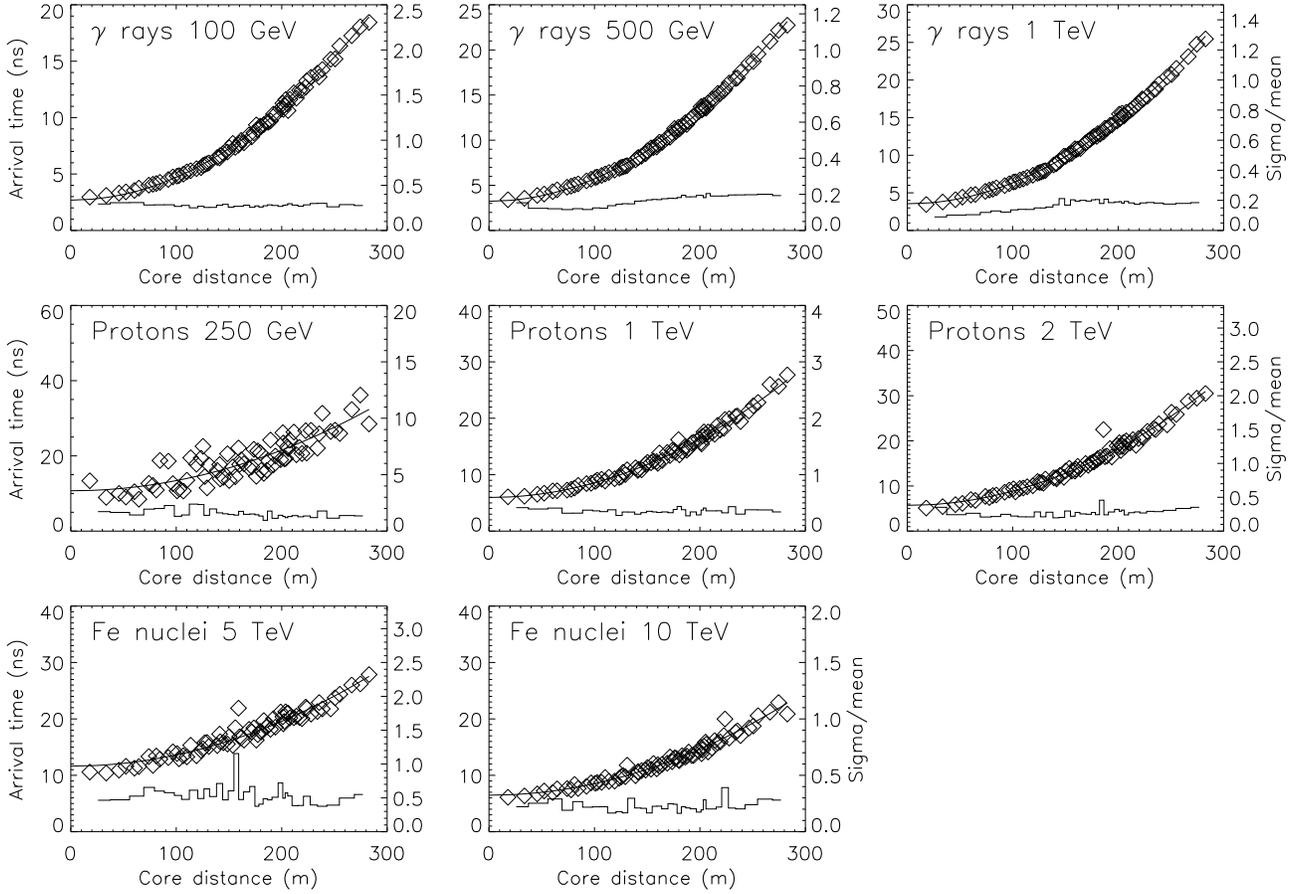,height=12cm}}
\caption{Variation of the mean arrival time of \v Cerenkov shower front
with the core distance for $\gamma -$ rays, protons and Fe nuclei of various 
energies as shown, averaged over several showers. Data from 5 consecutive 
detectors 
are averaged for plotting. Smooth curve corresponds to the best fit spherical
wavefront. Also shown in each plot are the relative shower to shower 
fluctuations (sigma/mean) of arrival times as a function of core distance.}
\end{figure}

\begin{table}
\caption{Radii of curvature from a spherical wavefront fit to \v Cerenkov
photon arrival times averaged over several (specified in text) showers. 
The radii are measured with respect to the observation level of Pachmarhi.}
\vskip 0.5cm
\begin{tabular}{lll}
\hline
Type of  & Energy of & Fitted radius  \\
primary  & the primary   & of curvature \\
         & ({\it GeV})     & ({\it m}) \\
\hline
$\gamma -$ rays & 100 & 8208 $\pm$ 44 \\
                & 500 & 6689 $\pm$ 37 \\
                & 1000 & 5981 $\pm$ 41 \\
protons & 250 & 6164 $\pm$ 125 \\
        & 1000 & 6401 $\pm$ 130 \\
        & 2000 & 5281 $\pm$ 155 \\
Fe nuclei & 5000 & 8383 $\pm$ 530 \\
          & 10000 & 8158 $\pm$ 395 \\
\hline 
\\
\end{tabular}
\end{table}

At lower energies proton showers exhibit more fluctuations in arrival
time than $\gamma -$ ray showers. Fig. 1 also shows shower to shower
fluctuations in the mean arrival time for $\gamma -$ rays,  protons and Fe 
nuclei as a function of core distance in terms of rms over mean.

\subsection{Altitude dependence}

\begin{table}
\caption{Radii of curvature from a spherical wavefront fit to \v Cerenkov
photons arrival times at observation levels of 2.2 km and sea level averaged
over typically 10 showers. The
radii are measured with respect to the observation level. It may be noted
that in order to get the height of shower maximum with respect to sea level 
one has to add the altitude of observation to the radii listed below.}

\vskip 0.5cm
\begin{tabular}{llll}
\hline
Altitude & Type of  & Primary & Fitted radius \\
above mean & primary  & Energy   & of curvature \\
sea level ({\it km})  & &({\it GeV}) & ({\it m}) \\
\hline
2.2 & $\gamma -$ rays & 500 & 5716 $\pm$ 32 \\

    & protons & 1000 & 5957 $\pm$ 135 \\

0.0 & $\gamma -$ rays & 500 & 7652 $\pm$ 47 \\

    & protons & 1000 & 8852 $\pm$ 325 \\

\hline
\\
\end{tabular}
\end{table}

A similar fitting procedure was adapted for delay profiles at two other 
observation levels {\it viz.} sea level and $2.2~km$ above mean sea level. The 
energies of the $\gamma -$ray and proton primaries are 500 and 1000 GeV 
respectively. Typically 10 showers were generated in each case. The
fitted radii are listed in table 2.

By a comparison of the fitted radii in tables 1 \& 2 it can be seen that the
fitted radii for a given species are independent of the observation altitude
when measured with respect to a common reference level.
Even though at a higher altitude the shower maximum is closer to the observation
level 
than at sea level, the bulk of the \v Cerenkov
photons are still emitted at the shower maximum. In other words, the radius
of curvature of the \v Cerenkov front is a measure of the majority
property while those photons produced elsewhere do not contribute
significantly to the curvature.

\section{Analysis of Pulse profiles}

\subsection{Pulse shape parameters}

Figure 2a \& 2b show the radial dependence of the mean arrival times for a
single $\gamma$ -ray shower of energy 500 GeV and a proton shower of
energy 1 TeV. The mean arrival time is computed over all the photons
arriving at each detector. Also shown in the figures are the RMS and the ratio
of RMS to mean arrival times for each detector.  It may be noted that the RMS 
fluctuations here refer to intra-shower fluctuations in photon arrival times. 
It can be readily seen
from a qualitative comparison that the distribution for proton primaries
is more fuzzy as the thickness of the shower front is larger at almost all
core distances. These plots demonstrate the existence of intra-shower
fluctuations in arrival times. It is intutively obvious that the shower
front characteristics have the signature of the primary species. In order
to investigate this we need to compute the \v Cerenkov pulse profiles as a
function of core distance and parameterize the pulse shapes. In doing so
one can choose such parameters like the rise time, decay time and full
width at half maximum (FWHM) which could be experimentally measured and
are physically meaningful. The rise time for example, reflects the
longitudinal growth of the cascade in the atmosphere while the decay time
exhibits the cascade attenuation past the shower maximum where as the FWHM is
a measure of the \v Cerenkov photon production profile [15,16].  Hence it is
important to understand and identify species dependent behavior of the
pulse shape parameters which could possibly be used in improving the
signal to noise ratio of the recorded \v Cerenkov signal.

\begin{figure}
\centerline{\psfig{file=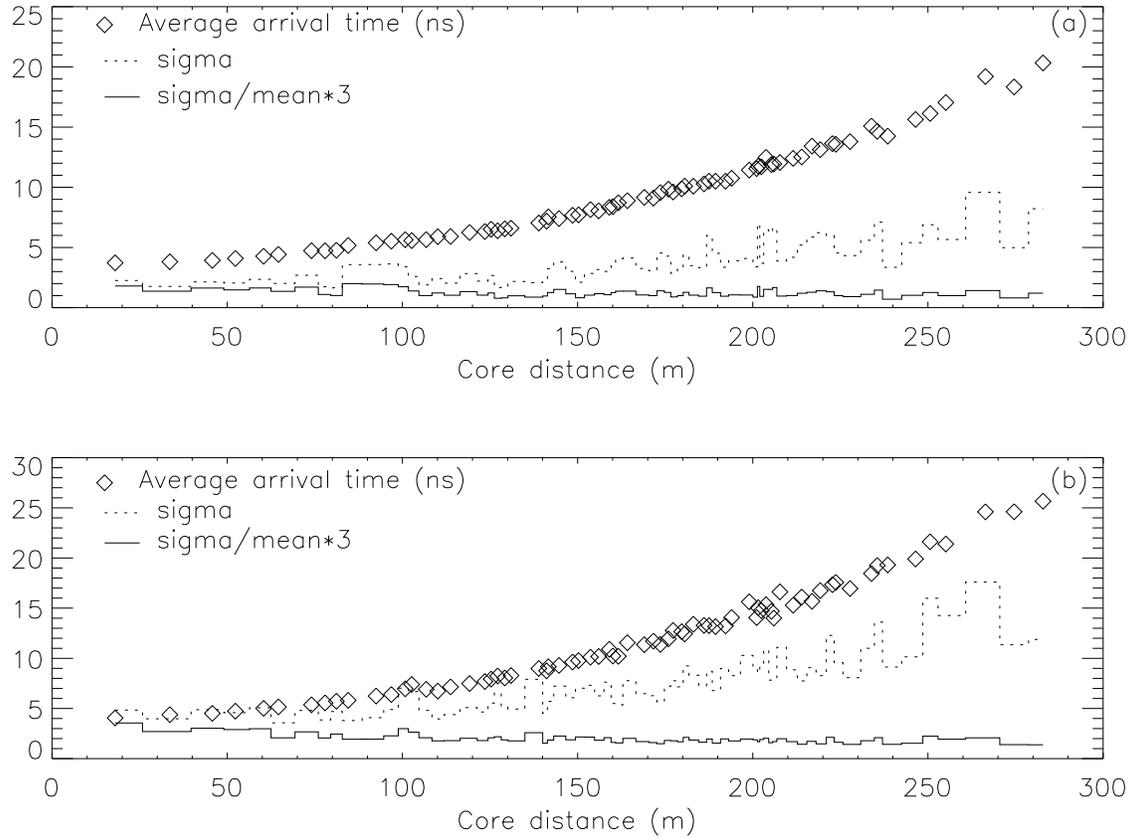,height=12cm}}
\caption{Figure showing the variation of average arrival time of \v
Cerenkov photons as a function of core distance for a single shower of
$\gamma -$ ray of energy 500 GeV $(a)$ and proton of energy 1 TeV $(b)$.
Average arrival time is computed for each detector and averaged over five
consecutive detectors for plotting. Also shown in the figure are the 
arrival time fluctuations measured in terms of RMS
of arrival times for each detector and the relative fluctuations expressed as a
ratio of the RMS to mean arrival time.}
\end{figure} 

With this in mind we carry out functional fits to the pulse shape profiles
using well known statistical probability density functions.

\subsection{The arrival time probability density function}

It is interesting to study the shape of the \v Cerenkov photon arrival
time delay distribution represented by the pulse shape at various core 
distances. It is convenient to
study the radial variation of the pulse shape by fitting a suitable function
like a probability density distribution function (PDF) to it. By doing so one 
can study
the behaviour of the parameters of the fitting functions directly or
better still the behaviour of experimentally measurable pulse shape
parameters which could be defined in terms of the function parameters.

We have tried to parameterize the pulse shapes by fitting well known PDF's
through multiparameter curve-fitting technique. Two types of functions
have been tried. 

\subsubsection{Fit to a $\Gamma$ - function}

The temporal \v Cerenkov profiles are described by a gamma
function of the form:

\begin{equation}
f(t) = c \times t^{(b-1)} exp(-a.t)
\end{equation}

where {\it t} is the arrival time of \v Cerenkov photons at the observation
plane, while {\it a} and {\it b} are parameters expressed in terms of 
$t_{max}$, time at the pulse maximum and the mean arrival time $<t>$. The
best fit parameters are derived by minimizing the $\chi^2$ for each
shower. Functional fits are also carried out using mean and variance as the
variables (Appendix A) without any change in conclusions. The fitted parameters and the corresponding reduced $\chi^2$ for
three different types of primaries at three different core distances are
listed in table 3.

\subsubsection{Fit to a lognormal distribution function}

We have also fitted a log-normal function for the arrival time
distribution of \v Cerenkov photons whose functional form is:

\begin{equation}
f(t) = {{1} \over {t \sigma \sqrt{2\pi}}} exp{-{1} \over {2\sigma^2}} (log{t}-\mu )^2 
\end{equation}

where $\mu$ and $\sigma$ are the mean and the standard deviation
respectively of the log-normal distribution. The fitted parameters and the
corresponding reduced $\chi^2$ for three different types of primaries at
three different core distances are listed in table 4. It was found that
this function could provide a better approximation of the actual delay
distribution with a long tail due to photons arriving with large delays.

Fits to the \v Cerenkov photon arrival time distributions from relatively
low energy photon and hadronic primaries to both these functions are
carried out to assess their relative merits in describing the time
profiles. Figure 3 shows the results of the fits for a typical
shower generated by a $\gamma $-ray of energy 500 GeV, a proton of energy
1 TeV and 10 TeV Fe nucleus. As we will see later the
profiles are also a function of core distance. Hence in each case, three
detectors, each corresponding to pre-hump, hump and post-hump region are
selected. Data are shown as a histogram binned over 0.25 ns bin for
$\gamma -$ ray and 0.5 ns for proton \& Fe primaries. The
functional fits are represented by smooth curves. Starting parameters for
fitting were derived from the data (see appendices A and B), and the best
fit parameters are obtained by $\chi^2$ minimization. The best fit
parameters and corresponding values of reduced $\chi^2$'s are listed in
Tables 3 and 4 for $\Gamma -$ and log-normal functional fits respectively.

\begin{figure}
\centerline{\psfig{file=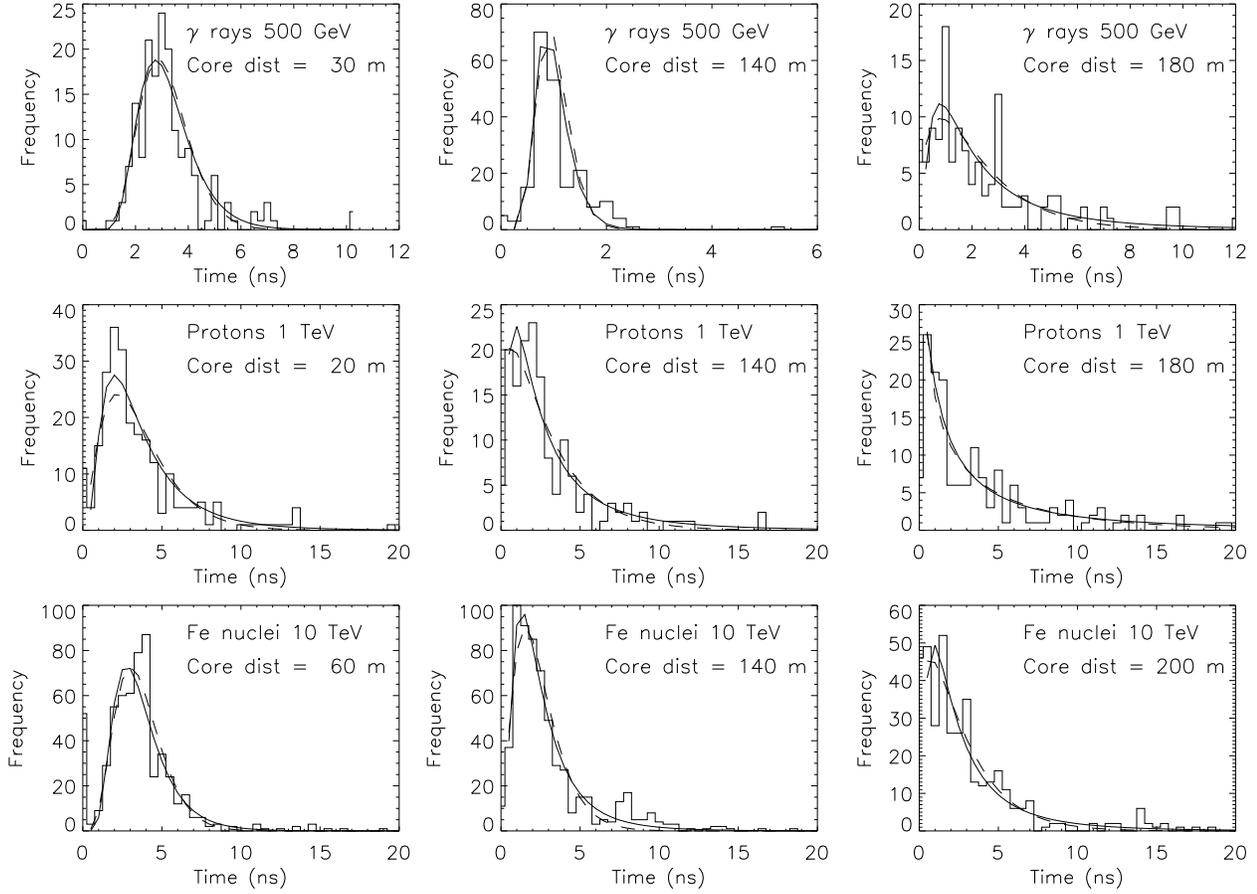,height=12cm}}
\caption{ Arrival time distribution of \v Cerenkov photons generated by a
500 GeV $\gamma -$ray, 1 TeV proton and 10 TeV Fe nucleus as incident on
detectors in pre-hump,  hump and post-hump regions, indicated by radial 
distances. Lognormal and $\Gamma -$ function fits to the distribution are 
shown by continuous and broken lines respectively.} 
\end{figure}

\begin{table}
\caption{Fitted pulse shape parameters from a $\Gamma -$ function}
\vskip 0.5cm
\begin{tabular}{lllllll}
\hline
Species & energy & core distance & a & b &  $\chi^2$ \\
        & (TeV)  &(m) & (ns)  & (ns)     &          \\
\hline
$\gamma -$ ray & 0.5 & 32 & 3.38 & 10.6 & 1.8 \\
               &     & 140 & 10.6 & 10.7 & 2.5 \\
               &     & 182 & 0.69 & 1.56 & 1.3  \\
proton         & 1   & 20 & 0.72 & 2.56 & 1.7 \\
               &     & 139 & 0.44 & 1.28 & 0.87 \\
               &     & 180 & 0.13 & 0.52 & 1.2 \\
Fe nuclei      & 10  & 64 & 1.56 & 5.72 & 0.76 \\
               &     & 141 & 1.09 & 2.66 & 3.0 \\
               &     & 203 & 0.49 & 1.34 & 1.38 \\
\hline
\end{tabular}
\end{table}
  
\begin{table}
\caption{Fitted pulse shape parameters from a log-normal distribution function}
\vskip 0.5cm
\begin{tabular}{llllll}
\hline
Species & energy &core distance & $\mu$ & $\sigma$  & $\chi^2$ \\
        &  (TeV) &(m) & (ns)  & (ns)     &           \\
\hline
$\gamma -$ ray & 0.5 & 32 & 1.12 & 0.11 & 1.6 \\
               &     & 140 & -0.05 & 0.099 & 1.7 \\
               &     & 182 & 0.67 & 0.90 & 1.3  \\
proton         & 1   & 20 & 1.15 & 0.46 & 1.2 \\
               &     & 139 & 0.84 & 0.98 & 0.81 \\
               &     & 180 & 1.097 & 2.55 & 1.2 \\
Fe nuclei      & 10  & 64 & 1.221 & 0.205 & 1.07 \\
               &     & 141 & 0.773 & 0.509 & 1.81 \\
               &     & 203 & 0.790 & 0.889 & 1.52 \\
\hline
\end{tabular}
\end{table}
 
We find that for both $\gamma -$ rays and hadronic primaries, lognormal
distribution provides a marginally better fit (average $\chi^2$ =1.35) for
the \v Cerenkov pulse profile at all core distances than the $\Gamma $-
function (average $\chi^2$ =1.6). Also it can be seen from appendix B that in
case of lognormal function, the observables like rise time, decay time amd
FWHM can be easily expressed in terms of the mean arrival time and the
RMS. This function could also reflect some intrinsic feature of the
underlying processes as noted by Battistoni {\it et al.}[11] that log-normal
distribution arises whenever the variable under study whose value takes a
random proportion of that of the previous step in a stochastic process
[17]. Here after we fit pulse profiles from all
the detectors with lognormal distribution function (LDF) and study the
variation of pulse shape parameters as a function of core distance and also as
a function of primary energy.

\begin{figure}
\centerline{\psfig{file=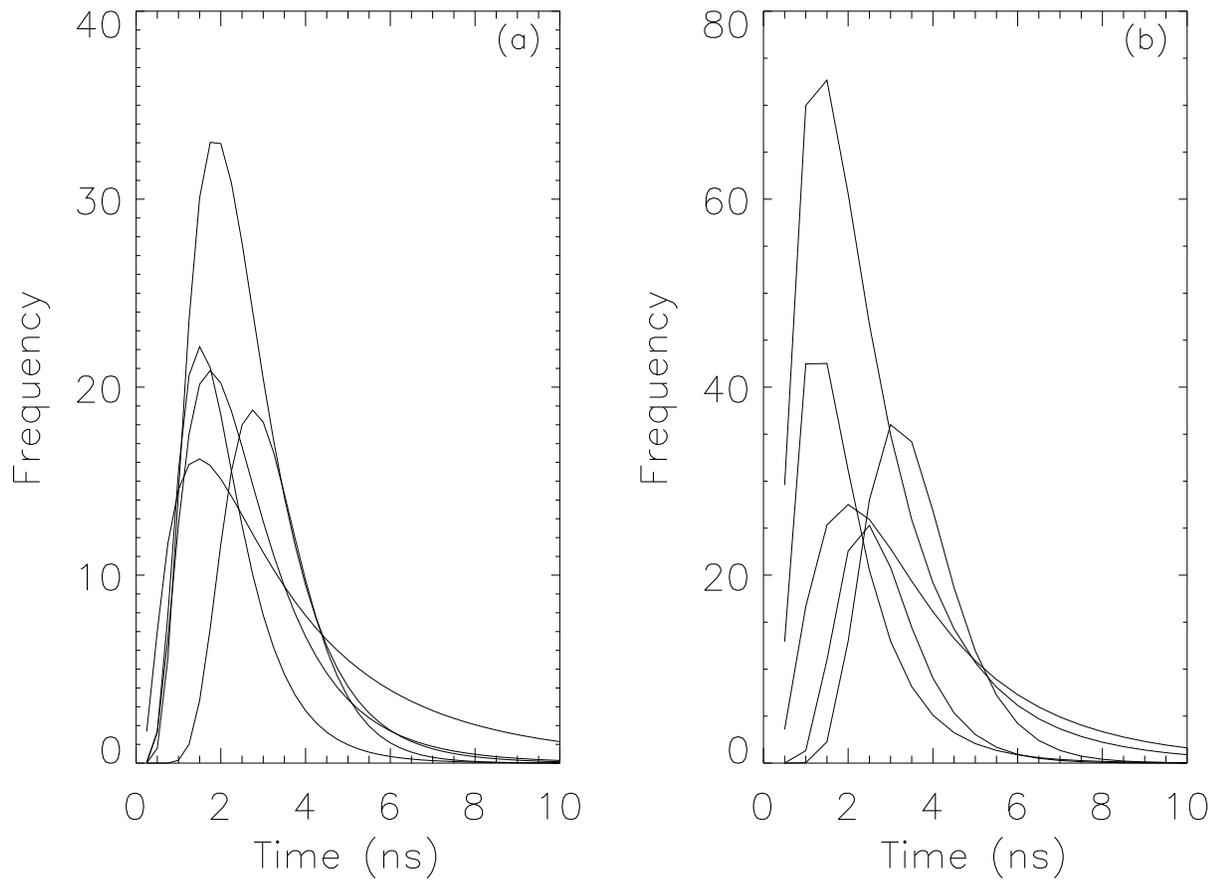,height=12cm}}
\caption{Lognormal fits to arrival time distributions of \v Cerenkov photons 
generated by 5 showers of  
(a) 500 GeV $\gamma -$rays and (b) 1 TeV protons as incident on a
detector in the pre-hump region.} 
\end{figure}

The sharp precursor pulse seen prominently in the case of Fe primary (pre-
hump region) is presumably due to \v Cerenkov light produced by muons. It has
been noted that these photons are generated at low altitudes ($1-3~kms$) with
a very small spread in production height suggesting that they are produced by
very few muons. The time separations between the $\mu$ and $e$
generated pulses and the pulse shape of the former are comparable with that 
reported by Cabot $et~al.$[18] as well as
Roberts {\it et al.}[19].

Fig. 4
shows similar fits to \v Cerenkov pulses in a detector in the pre-hump region
from a set of showers generated by (a) 500 GeV $\gamma -$rays 
and (b) 1 TeV
protons. The diversity in the pulse shapes seen here demonstrate the effect of 
fluctuations in photon number as well as pulse shape parameters.

\section{Behaviour of Pulse shape parameters}

\subsection{Core distance dependence}

For each shower, the pulse profile from each detector was fitted with a
LDF as mentioned before. For each detector, the arithmetic mean and the
variance were calculated from the data set consisting of individual
arrival times of detected \v Cerenkov photons by a detector at a given core 
distance. The LDF parameters $\mu$
and $\sigma$ are then derived using the mean and RMS from the data, 
as described in appendix B.  Using these
derived values of $\mu$ and $\sigma$ as initial guesses,  pulse
profile, which is suitably binned, is fitted.  Using the expressions given 
in appendix B, pulse shape
parameters, which include rise time, decay time and FWHM, are calculated
both for derived and fitted LDF parameters. It may be noted that the pulse
shape parameters calculated directly from the data (which are sometimes
referred to as predicted parameters) are free from systematic errors
arising out of data binning.

\subsubsection{$\gamma -$ray and proton primaries}

      Fig. 5a shows the variation of mean rise time (duration for the pulse
to reach from 10\% to 90\% of maximum height), as a function of core
distance generated by  $\gamma -$ rays of energy 500 GeV. Average is taken
over 10 showers. Rise time from the predicted LDF is shown with a
continuous line and that from the fitted LDF is shown by diamonds. Shower
to shower fluctuations for predicted and fitted rise time are shown in fig. 5b. 
expressed in terms of RMS/mean. There is a good agreement
between the predicted and fitted rise times at all core distances,
indicating that in practice there is no need to fit the pulse profiles
explicitly.

\begin{figure}
\centerline{\psfig{file=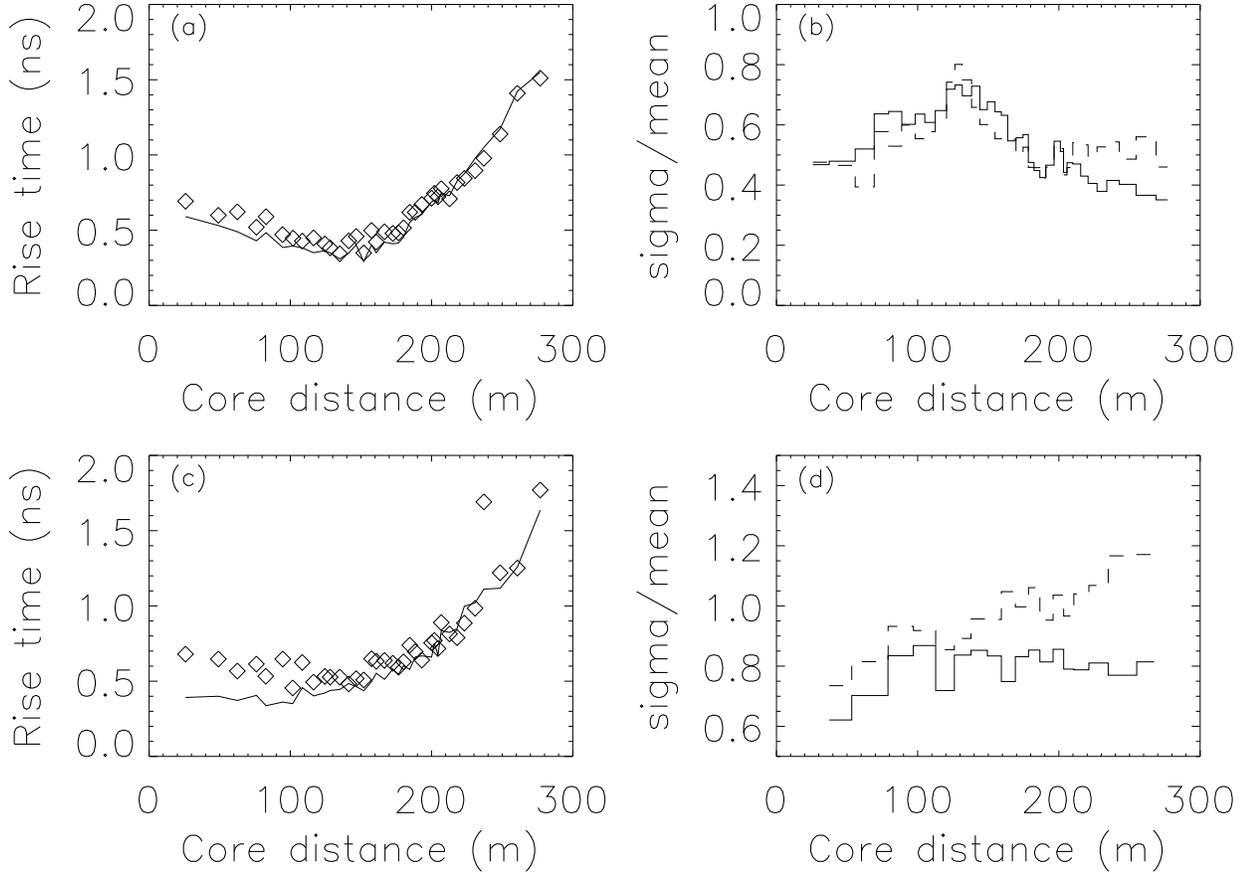,height=12cm}}
\caption{$(a)$ Radial variation of mean rise time of a \v Cerenkov pulse,
generated by $\gamma-$ ray primaries of energy 500 GeV and $(c)$ 1 TeV protons
incident vertically at the top of the atmosphere. Average is taken
over 10 showers for each detector. Rise time from the predicted LDF is
shown as a continuous line, whereas that from the fitted LDF is shown with
diamonds. $(b)$ Variation of shower to shower fluctuations of rise time in
terms of RMS/mean as a function of core distance for predicted (continuous
line) and fitted distribution functions (broken line) for $\gamma -$ ray 
primaries and $(d)$ for proton primaries.}
\end{figure}

     Fig. 5c shows core distance dependence of rise time for proton
primaries of energy 1 TeV. Again there seems to be a good agreement between
predicted and fitted rise times. While the rise times {\it per se} are not 
species dependent, shower to shower fluctuations in proton
showers, as shown in fig. 5d, are larger compared to the corresponding
fluctuations for $\gamma -$ rays by about a factor of 1.5 - 2.0. 

\begin{figure}
\centerline{\psfig{file=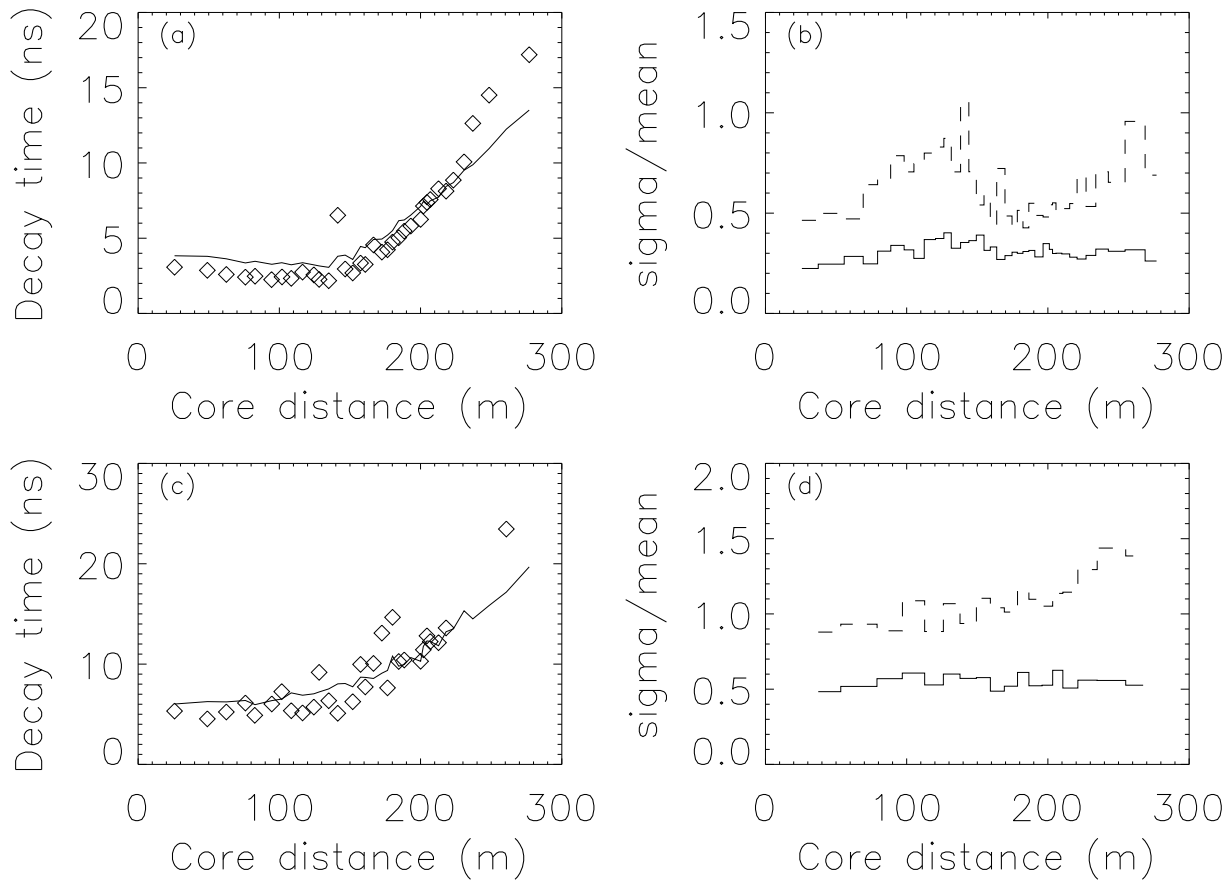,height=12cm}}
\caption{$(a)$ Radial variation of mean decay time of a \v Cerenkov pulse,
generated by $\gamma-$ ray primaries of energy 500 GeV and $(c)$ 1 TeV protons 
incident vertically on the top of atmosphere. Average is taken
over 10 showers for each detector. Decay time from the predicted LDF is
shown as a continuous line, whereas that from the fitted LDF is shown with
diamonds. $(b)$ Variation of shower to shower fluctuations of decay time in
terms of RMS/mean as a function of core distance for predicted (continuous
line) and fitted distribution function (broken line) for $\gamma -$ray 
primaries and $(d)$ for proton primaries.}
\end{figure}

     Variation of decay time (defined as the time duration during which
pulse falls from 90\% to 10\% of maximum height) with core distance for
$\gamma -$ ray and proton primaries is shown in figs. 6a and 6c respectively. 
Decay times from proton primaries are marginally longer than those from
$\gamma -$ray primaries.
Also shown in the same figure (6b and 6d respectively) are the relative
shower to shower fluctuations in decay times. Once again the proton
primaries exhibit larger fluctuations compared to those from $\gamma -$ray
primaries. In both the cases agreement between the predicted and fitted
decay times is poorer at larger core distances mainly because of the
paucity of photons and uncertainties due to binning. Shower to shower 
fluctuations too contribute to this discrepancy.

\begin{figure}
\centerline{\psfig{file=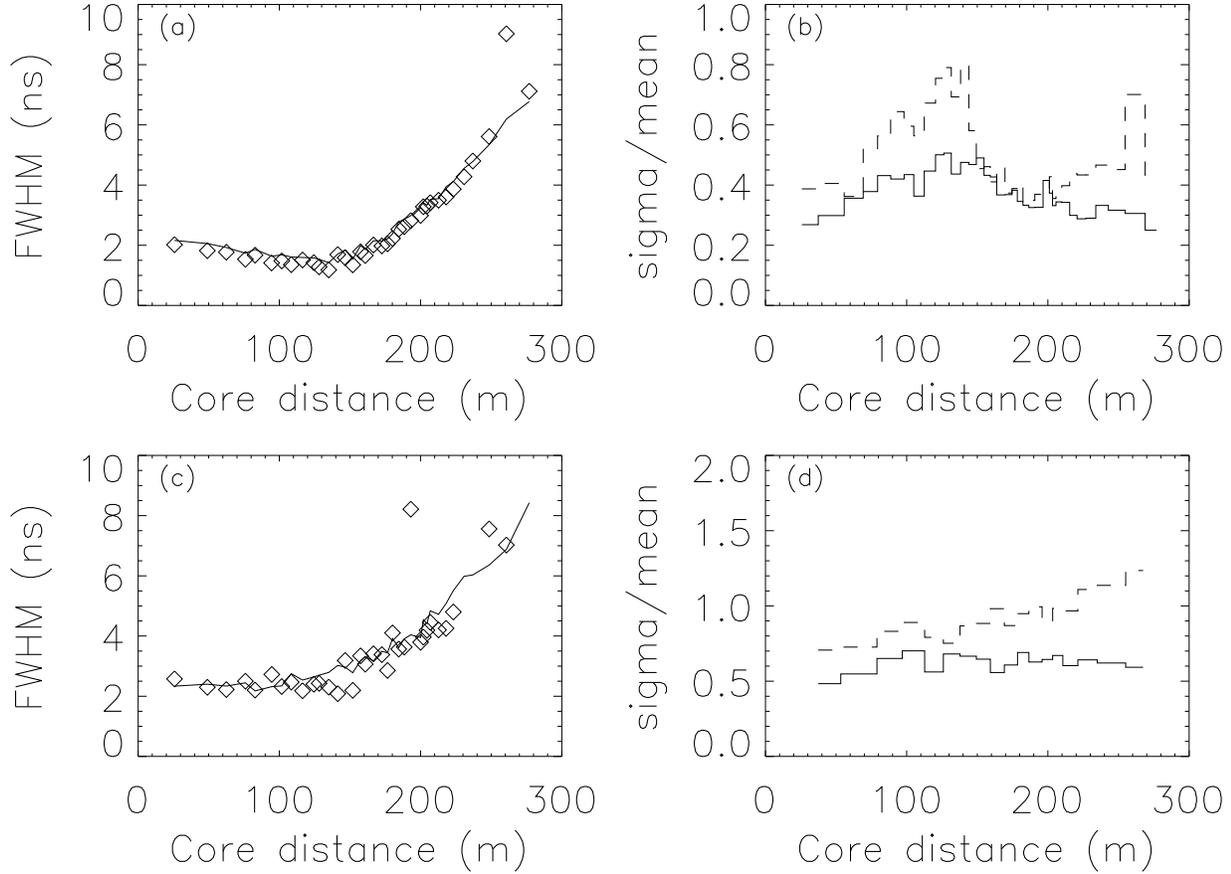,height=12cm}}
\caption{$(a)$ Radial variation of mean FWHM of a \v Cerenkov pulse,
generated by $\gamma-$ ray primaries of energy 500 GeV and $(c)$ 1 TeV protons
incident vertically at the top of atmosphere. Average is taken
over 10 showers for each detector. FWHM from the predicted LDF is shown as
a continuous line, whereas that from the fitted LDF is shown with
diamonds. $(b)$ Variation of shower to shower fluctuations of FWHM in terms
of RMS/mean as a function of core distance for predicted (continuous line)
and fitted distribution function (broken line) for $\gamma -$ray primaries and
$(d)$ proton primaries.}
\end{figure}

     Radial dependence of FWHM for $\gamma -$ ray and proton primaries is
shown in figs. 7a and 7c, respectively. A general agreement is seen
between the predicted and fitted FWHM. Also shown in the same figure (7b \& 7d 
respectively) are the
relative shower to shower fluctuations in FWHM for $\gamma -$ray and
proton primaries. Once again the relative fluctuations are about a factor
of 2 higher in the case of latter species.  For all the pulse shape
parameters discussed above, $\gamma -$ ray showers show a better agreement
between the fit and prediction compared to protons because of reasons mentioned
before. 

The radial variations of the relative errors on the pulse shape parameters as
shown in figs. 5b, 5d, 6b, 6d, 7b \& 7d are not greatly influenced by the 
number of showers used. They have, however, rather large errors of 47\%, 26\% 
and 30\% respectively for rise time, decay time and FWHM arising primarily 
from the small number of showers used here. The peaks seen in
the case of $\gamma -$ray primaries are more a result of the mean values 
showing a minimum at the position of the hump. 

\subsubsection{Fe primaries}

\begin{figure}
\centerline{\psfig{file=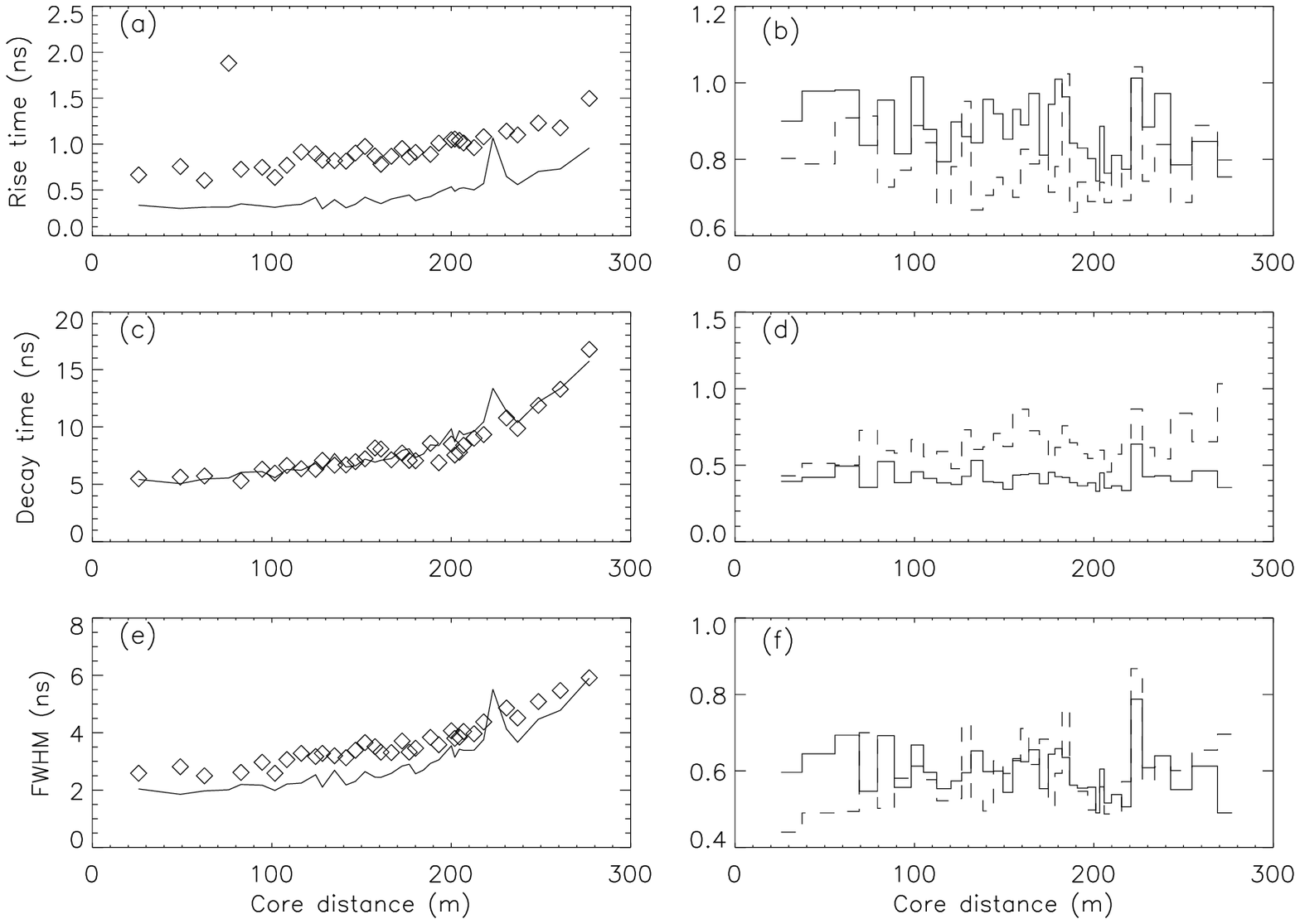,height=12cm}}
\caption{Radial variation of mean $(a)$ rise time, $(c)$ decay time and $(e)$ FWHM
of a \v Cerenkov pulse,
generated by primary Fe nuclei of energy 10 TeV. Average is taken over 10
showers for each detector. Rise time from the predicted LDF is shown as a
continuous line, whereas that from the fitted LDF is shown with diamonds.
Variation of shower to shower fluctuations of $(b)$ rise time, $(d)$ decay time
and $(f)$ FWHM in terms of
RMS/mean as a function of core distance for predicted (continuous line)
and fitted distribution function (broken line).}
\end{figure}

Figure 8a, 8c \& 8e show the radial variation of rise time, decay time
and FWHM respectively for a primary iron nucleus of energy 10 TeV incident
vertically at the top of the atmosphere. It can be seen that the predicted
rise time alone is consistently smaller than that obtained by fitting a
LDF while decay time and FWHM show a good agreement between the two. 
It was noticed that in  
the case of Fe primaries the pulse profile shows a rather long tail due to a
few \v Cerenkov photons arriving at large delays. The LDF is inadequate to fit
these few isolated stragglers and hence the fitted parameters are not affected.
The derived parameters, on the other hand, are affected by these
delayed photons.  This is also reflected in the fact that the LDF generated 
by the parameters derived from the data show a poor fit resulting in large 
$\chi ^2$ value while the fitted LDF shows a $\chi ^2$ around 1.0. 
In short, for Fe primaries, the derived values of decay time and FWHM do 
represent the true value while the rise time does not.

Similarly, figures 8b, 8d \& 8f show the relative shower to shower
fluctuations of rise time, decay time and FWHM respectively for Fe
primaries. As for proton primaries the relative fluctuations are generally
larger than those for $\gamma -$ray primaries and are almost independent of core
distance. The relative fluctuations computed for derived \& fitted
parameters are generally in agreement with each other in all the three
cases. 

\subsection{Primary species dependence}

The sensitivity of the pulse shape parameters to the primary species can
be judged by a comparison of their radial variation for $\gamma -$rays,
proton and Fe primaries as shown in figures 5-8. It may be seen that the
proton \& Fe primaries exhibit larger shower to shower fluctuations as
expected.  However the magnitude of pulse rise time shows no significant
difference for the two types of primaries except that the hadronic
primaries show a monotonic increase with core distance. On the other hand
pulse fall time and width show more perceptible differences. The
possibility of using these parameters for improving the $\gamma -$ray
signal strength will be dealt in a forthcoming paper.

\subsection{Energy dependence of pulse shape parameters}
 
      We have generated several showers of $\gamma -$ rays, with primary
energies in the range 0.5-10 TeV, following a powerlaw distribution with
index -2.65. Proton showers were also generated following the same power
law index, in the energy range 0.5-20 TeV.  For each shower, mean pulse
shape parameters were derived as described above and averaged over all the
detectors. Fig. 9 shows the variation of different pulse shape
parameters with primary energy. The sample consists of 34 showers of
$\gamma -$ rays and 70 showers of protons. Also shown in the figure are
the mean pulse shape parameters, averaged over a number of monoenergetic 
showers of $\gamma -$ rays and protons. For $\gamma -$ rays of energies
250 GeV, 500 GeV, 1 TeV and 3.5 TeV as well as for protons of energies
500 GeV, 1 TeV, 2 TeV and 5 TeV, mean was calculated over 10, 10, 9 and
4 showers respectively. At lower energies there seems to
be rather large fluctuations which are presumably due to the smaller
number of \v Cerenkov photons at lower primary energies. Supported by
points derived from monoenergetic showers there seems to be a weak linear
correlation between the shape parameters and primary energy.

\begin{figure}
\centerline{\psfig{file=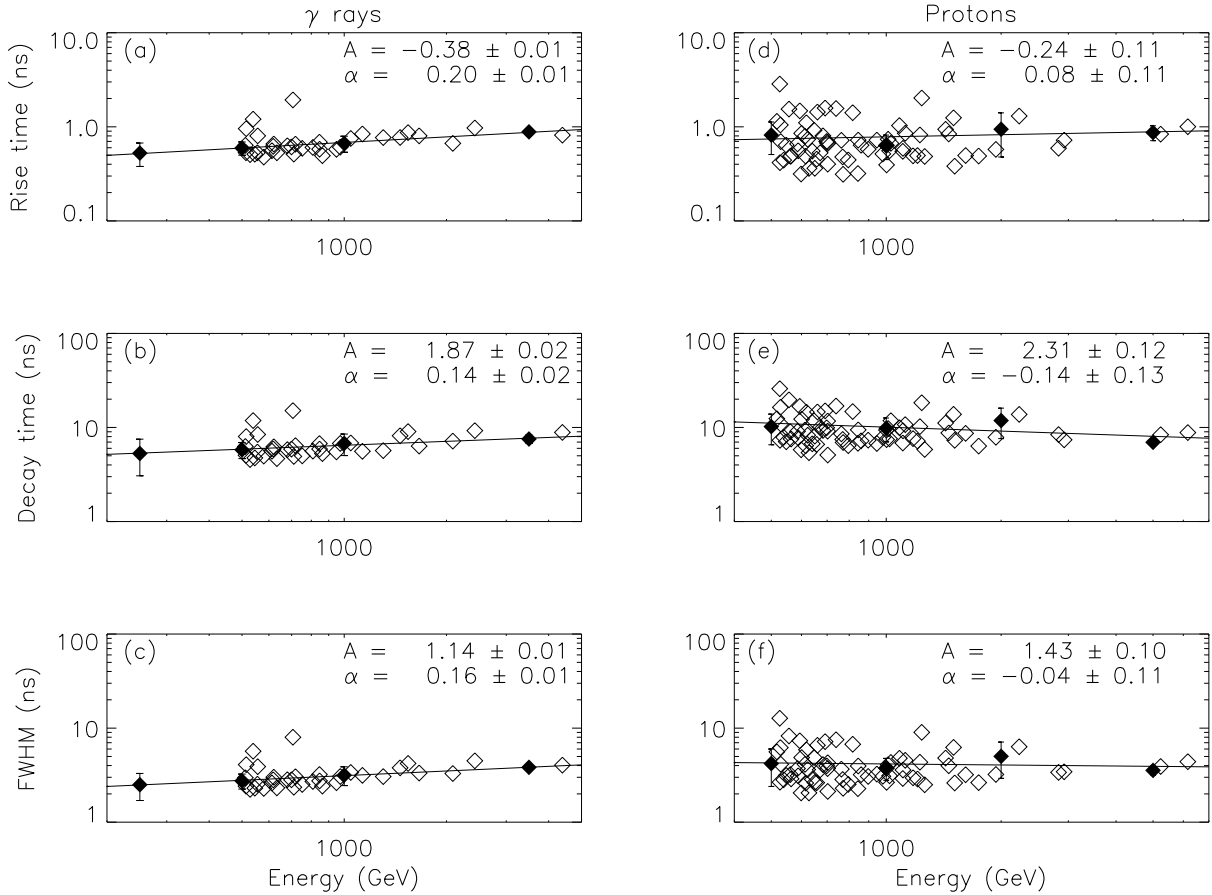,height=12cm}}
\caption{Variation of average $(a)$ rise time, $(b)$ decay time and $(c)$
FWHM of \v Cerenkov pulse with primary energy for $\gamma -$ rays. Similar
variation for proton showers is shown in $(d)$, $(e)$ and $(f)$, respectively.
Pulse shape parameters averaged over all the detectors for each shower are
shown by hollow diamonds. For $\gamma -$ rays of energy 250 GeV, 500 GeV,
1 TeV and 3.5 TeV and for protons of energy 500 GeV, 1 TeV, 2 TeV and 5
TeV, mean pulse shape parameters averaged over a number of showers are
shown with filled diamonds along with RMS error. For monoenergetic showers
variation of pulse shape parameters with energy are fitted with a relation
$AE^{-\alpha}$, where E is energy of the primary species. Values of A and
$\alpha$ are indicated in the figure for each parameter. }
\end{figure}

\subsection{Altitude dependence}

\begin{figure}
\centerline{\psfig{file=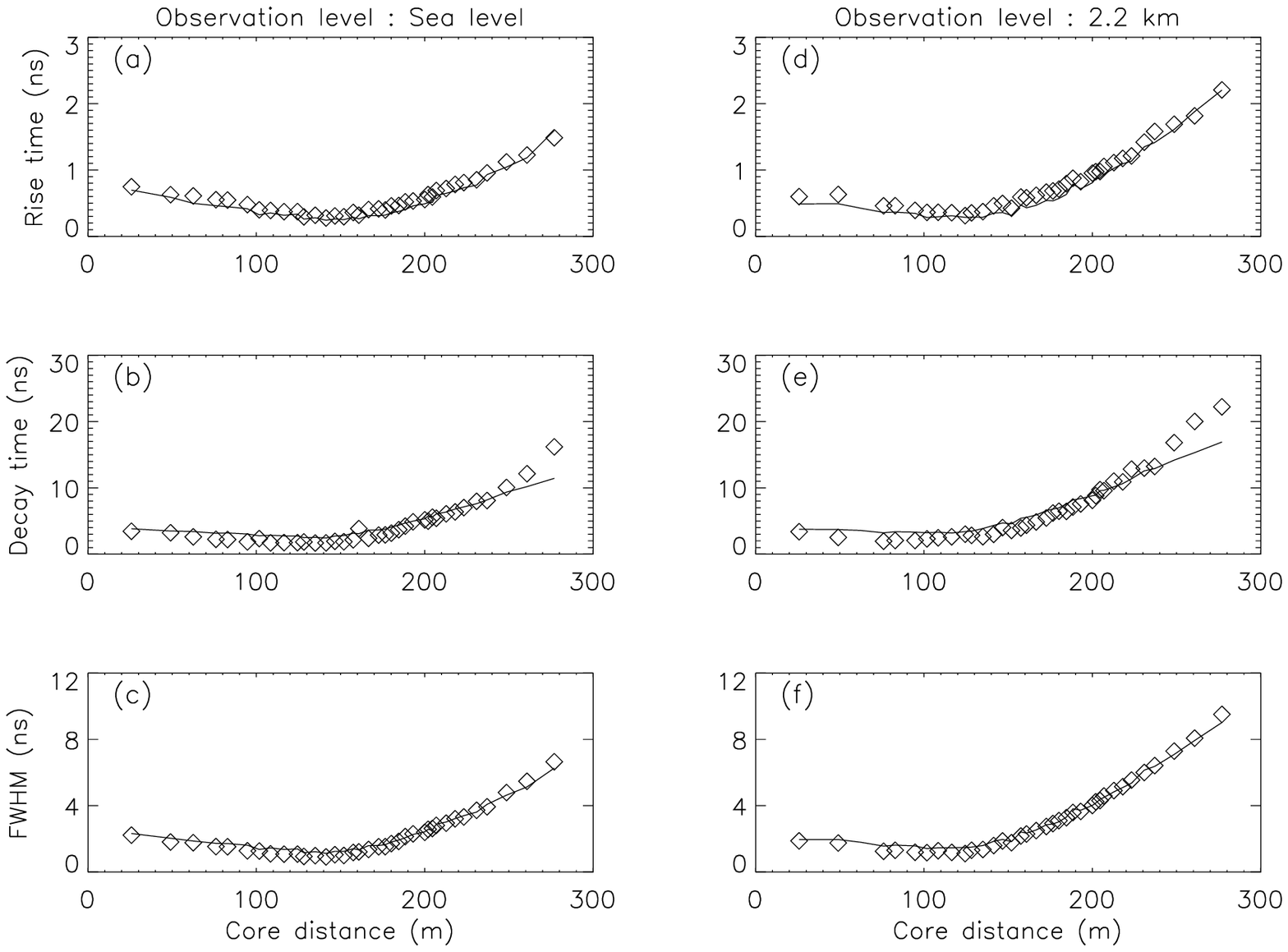,height=12cm}}
\caption{Radial variation of average $(a)$rise time, $(b)$ decay time and $(c)$
FWHM of a \v Cerenkov pulse as
observed at sea level and the same ($d$, $e$ and $f$ respectively) at an 
observation level of 2.2 km above
mean sea level from 500 GeV $\gamma -$ rays incident vertically at the top
of atmosphere. Pulse shape parameters derived from the LDF are shown as a 
continuous line while fitted values are shown as diamonds.}
\end{figure}

\begin{figure}
\centerline{\psfig{file=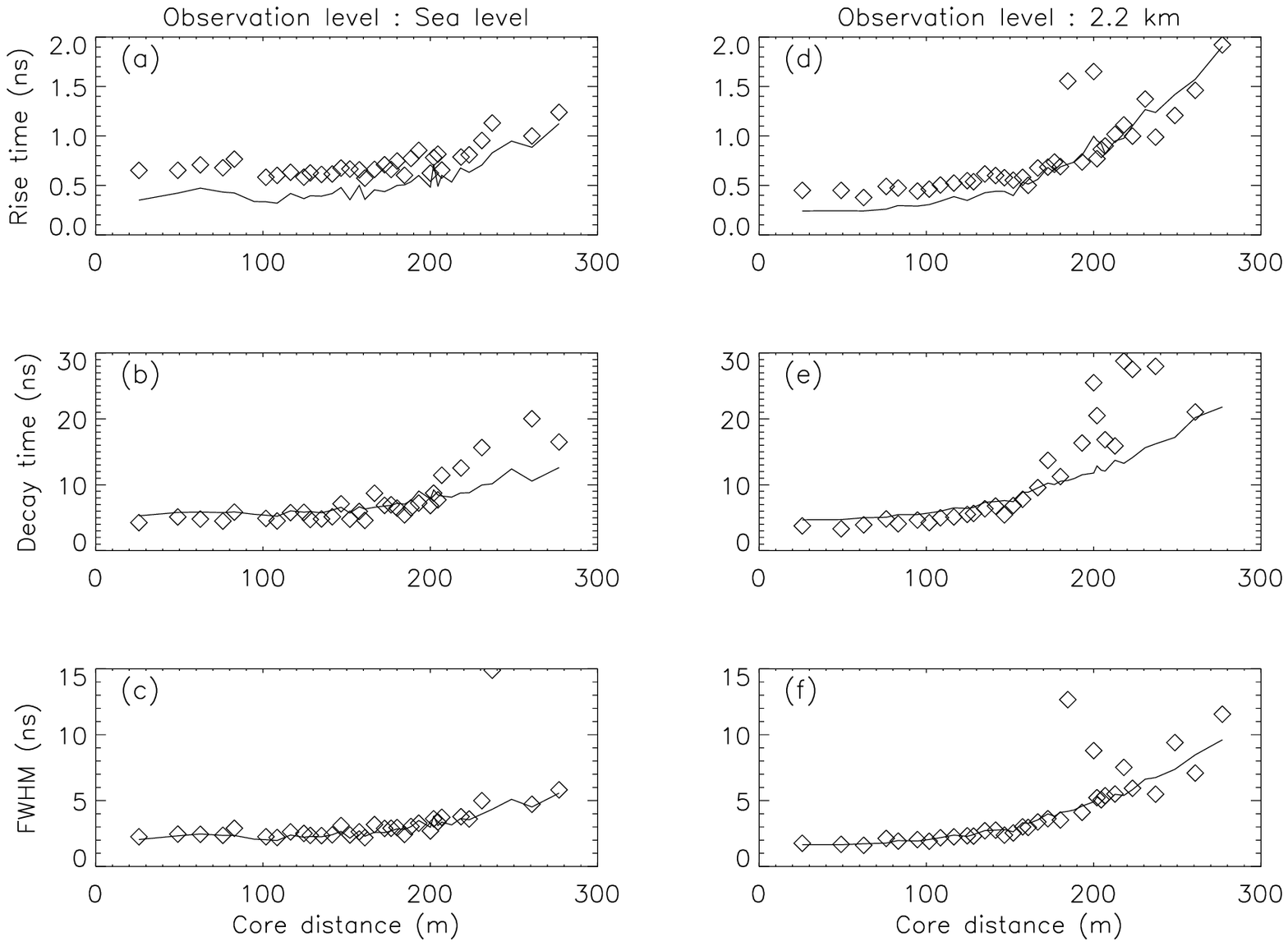,height=12cm}}
\caption{Radial variation of average $(a)$rise time, $(b)$ decay time and $(c)$
FWHM of a \v Cerenkov pulse as
observed at sea level and the same ($d$, $e$ and $f$ respectively) at an 
observation level of 2.2 km above
mean sea level from  1 TeV protons incident vertically at the top
of atmosphere.  Pulse shape parameters derived from the LDF are shown as a 
continuous
line while fitted values are shown as diamonds.}
\end{figure}

Radial variation of all the three pulse shape parameters were studied at
two other observation levels both for $\gamma -$ray and proton primaries.
Figure 10 shows the variation of rise time, decay time and FWHM
of the \v Cerenkov pulses as measured at various core
distances for $\gamma -$rays of energy 500 GeV incident vertically at the
top of atmosphere. The panels labeled {\it a, b} \& $c$ correspond to sea level
while those labeled {\it d, e} \& $f$ correspond to an observation level of 
2.2 km.
Similarly, Figure 11 shows the corresponding behavior for proton
primaries of energy 1 TeV.

It can be readily noticed that for $\gamma -$ray primaries there is
practically no change in the radial variation in the three parameters
until the hump region. However beyond the hump region some parameters seem
to vary faster at higher altitudes. The range of rise time, decay time and
FWHM at the three different observation levels are listed in table 5. It
can be seen from the table that both for $\gamma -$ray and proton primaries, 
the maximum values of all the three pulse shape parameters {\it viz.} rise time,
decay time \& FWHM at large core distances show a systematic increase with 
increase in observation level showing their possible sensitivity to altitude.
However assuming that a diverging \v Cerenkov cone is being intercepted at 
different observation levels, one would expect such an increase at higher 
altitudes due to simple geometric effects. Let us see if this indeed is the 
reason for this apparent increase. 

Since the linear increase in the parameter begins from the position of the hump,
it is obvious that the hump position plays an important role here. Therefore we 
estimated the core distance of the hump for a 500 GeV $\gamma -$ray primary to 
be 143 $m$, 130 
$m$ and 116 $m$ respectively at three altitudes sea-level, 1070 $m$ and 2200 $m$
above mean sea-level. These values are consistent with earlier estimates by Rao
\& Sinha [20]. It may be seen that at equal radial distances, measured 
from the hump position at a given level, the shape parameters have similar 
values at each altitude of observation. Thus, the shape parameters do not have 
intrinsic sensitivity to the observation altitude. The same seems to hold good 
for proton primaries as well even though their lateral distribution does not 
exhibit any prominent hump [14]. 

\begin{table}
\caption{Range of radial variation of rise time, decay time and the FWHM
of the \v Cerenkov pulse profile at various observation levels.}
\vskip 0.5cm
\begin{tabular}{lllllll}
\hline
Species & Energy & Observation       &           &            &  \\
        & GeV    & Level ({\it (km)} &           &            &  \\
        &        & Above mean        & Rise Time & Decay Time & FWHM \\
        &        & sea level)        & {\it ns}  & {\it ns}   & {\it ns} \\
\hline
$\gamma -$ rays & 500 & 0.0           & 0.35 - 1.5 & 2 - 16 & 1 - 6.5  \\
               &     & 1.0           & 0.35 - 1.5 & 2 - 17 & 1 - 7    \\
               &     & 2.2           & 0.3  - 2.2 & 2 - 24 & 1 - 10   \\
\hline
Protons        & 1000 & 0.0          & 0.3 - 1.35 & 4 - 21 & 2 - 6.5   \\
               &      & 1.0          & 0.35 - 1.8 & 5 - 23 & 2 - 8.5   \\
		   &      & 2.2          & 0.3 - 1.9  & 4 - 29 & 2 - 12    \\
		   
\hline
\\
\end{tabular}
\end{table}

\subsection{Incident angle dependence}

\begin{figure}
\centerline{\psfig{file=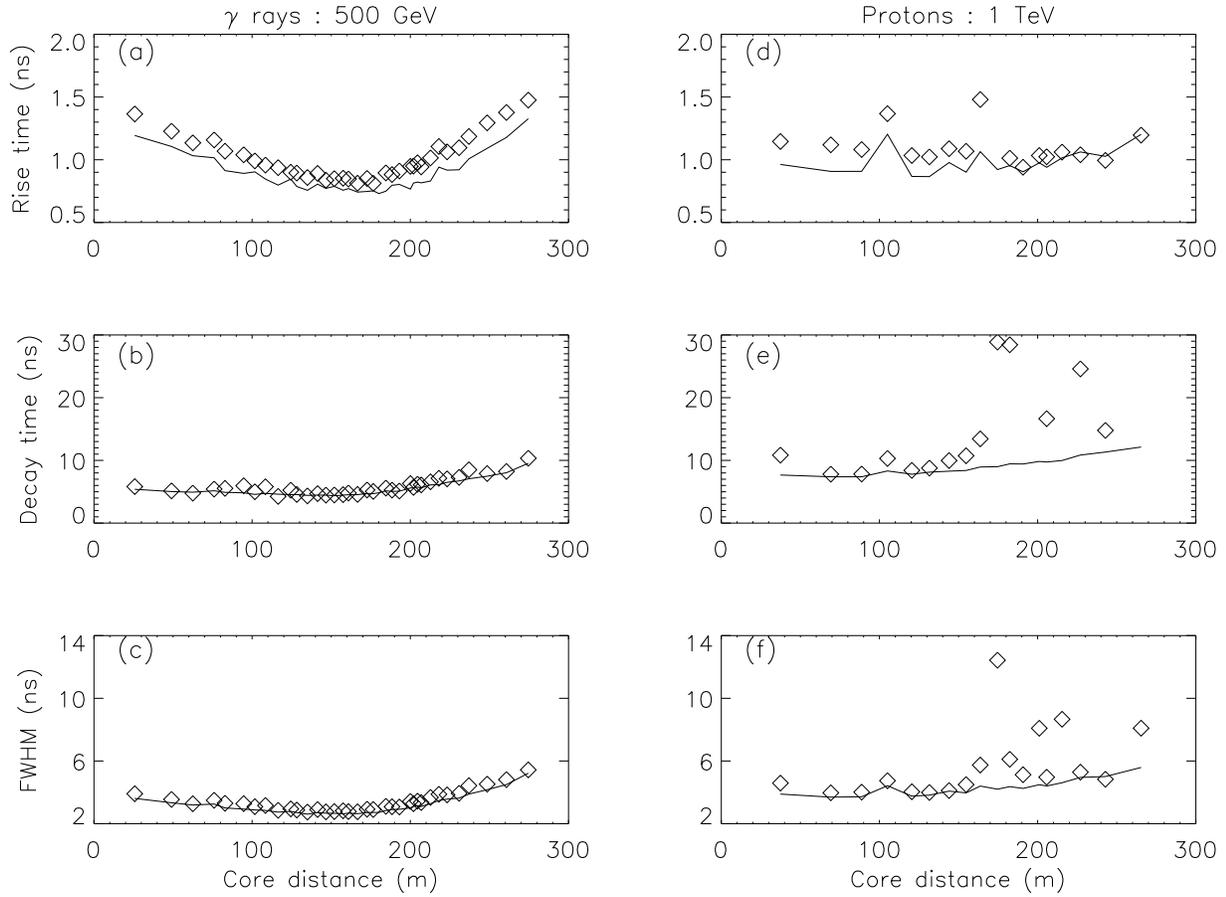,height=12cm}}
\caption{Radial variation of average $(a)$ rise time, $(b)$ decay time and $(c)$
FWHM of a \v Cerenkov pulse as
observed at Pachmarhi from  500 GeV $\gamma -$rays and the same quantities
($d$, $e$ and $f$ respectively) from 1 TeV
protons incident at an angle of $30^\circ $ with respect to vertical at
the top of atmosphere.  Pulse shape parameters derived from the LDF is shown as a
continuous line while the fitted values are shown as diamonds.}
\end{figure}

Figures 12a, 12b \& 12c show the radial variation of rise time, decay time
and the pulse width for 500 GeV $\gamma -$ray primaries incident at an
angle of 30$^\circ $ to the vertical at the top of atmosphere at
Pachmarhi.  Similarly figures $12d,~12e~\&~12f$ show the corresponding
variations for primary protons of energy 1 TeV. 

There are some noticeable differences between the \v Cerenkov pulse shape
characteristics of inclined showers compared to those of vertical showers
for both the types of primaries. Inclined $\gamma -$ray showers exhibit a
larger change in the pulse shape parameters in the core to hump region
even though the variation is comparable to the shower to shower
fluctuations as in the case of vertical showers. Radial variation of
relative shower to shower fluctuations on the other hand seems to be more
uniform and reduced in magnitude at larger incident angles as shown in
table 6. This could be a result of additional Coulomb scattering of
electrons due to increased path lengths for inclined showers. The hump
also moves away from the core by a $sec(\theta )$ factor.

Similarly, in the case of inclined proton showers the pulse shape
parameters have larger values near the core and remain more uniform as a
function of core distance compared to that for vertical showers.
Both for $\gamma -$ray and proton primaries which do not have significant 
muon content at the energies studied here the rise time distribution broadens
slightly with increasing zenith angle as expected from the increased distance
from the shower maximum [19].

\begin{table}
\caption{Range of radial variations of relative shower to shower
fluctuations in rise time, decay time and the FWHM of the \v Cerenkov
pulse profile for vertical and inclined showers.}
\vskip 0.5cm
\begin{tabular}{lllllll}
\hline
Species & Energy & Incident          & Rise Time & Decay Time & FWHM \\
        & GeV    & Angle ($^\circ$)  &           &            &      \\
\hline
$\gamma -$ rays & 500 & 0.0           & 0.35 - 0.75 & 0.25 - 0.4 & 0.25 -
0.5  \\
               &     & 30.0          & 0.24 - 0.4 & 0.17 -0.25 & 0.15 -
0.24 \\
\hline
Protons        & 1000 & 0.0          & 0.6 - 0.85 & 0.45 - 0.7 & 0.45 -
0.8  \\
               &      & 30.0         & 0.5 - 0.8 & 0.25 - 0.55 & 0.15 -
0.6  \\
		   
\hline
\\
\end{tabular}
\end{table}

\section{Discussions}

\subsection{The Spherical Shower front}

\begin{figure}
\centerline{\psfig{file=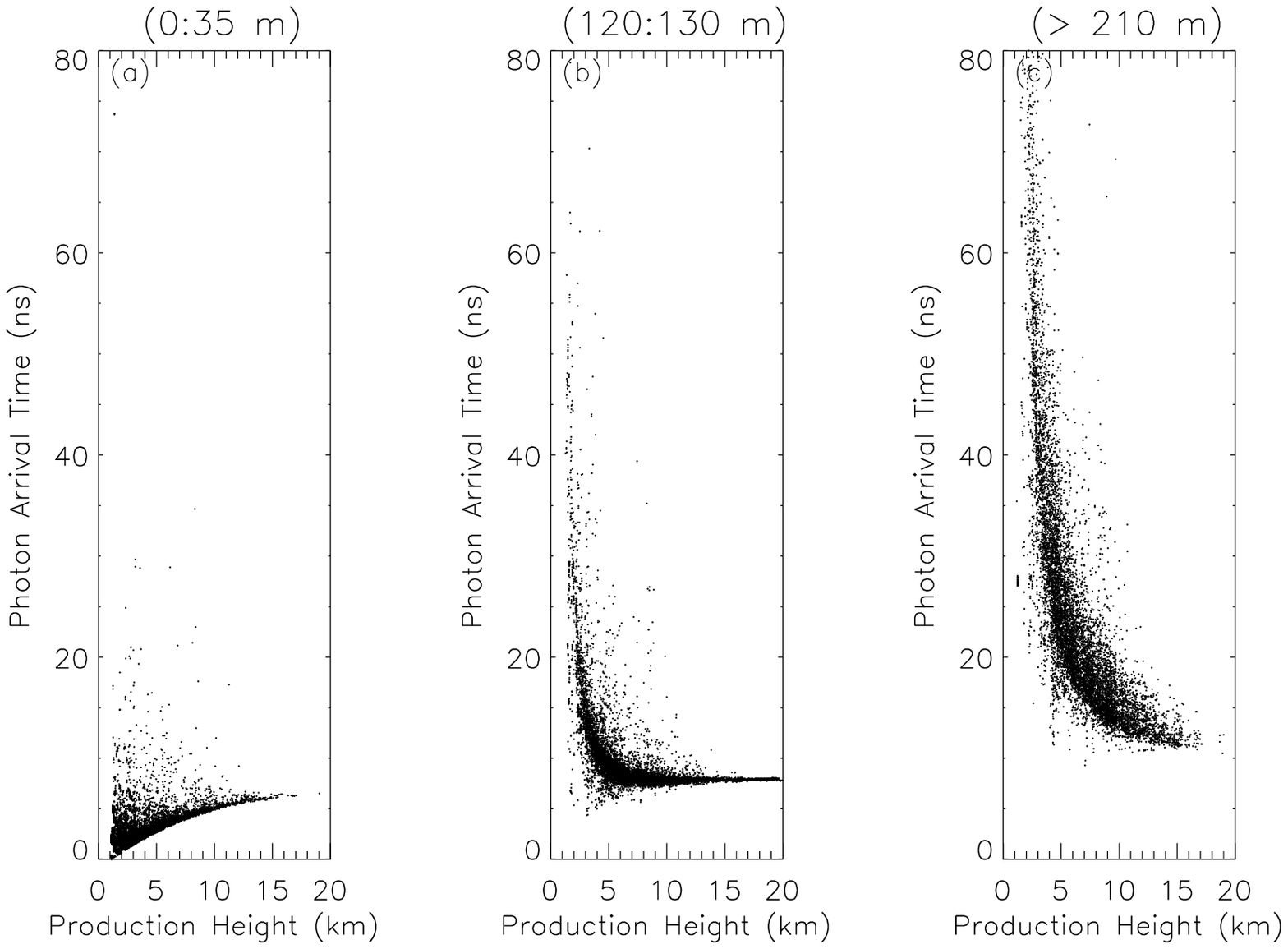,height=12cm}}
\caption{A plot of the \v Cerenkov photon arrival times as a function of 
production height, at three different core distance ranges: pre-hump, hump 
\& post hump regions. The primary is a gamma ray of energy 550 GeV incident 
vertically on the top of the atmosphere.} 
\end{figure}

\begin{figure}
\centerline{\psfig{file=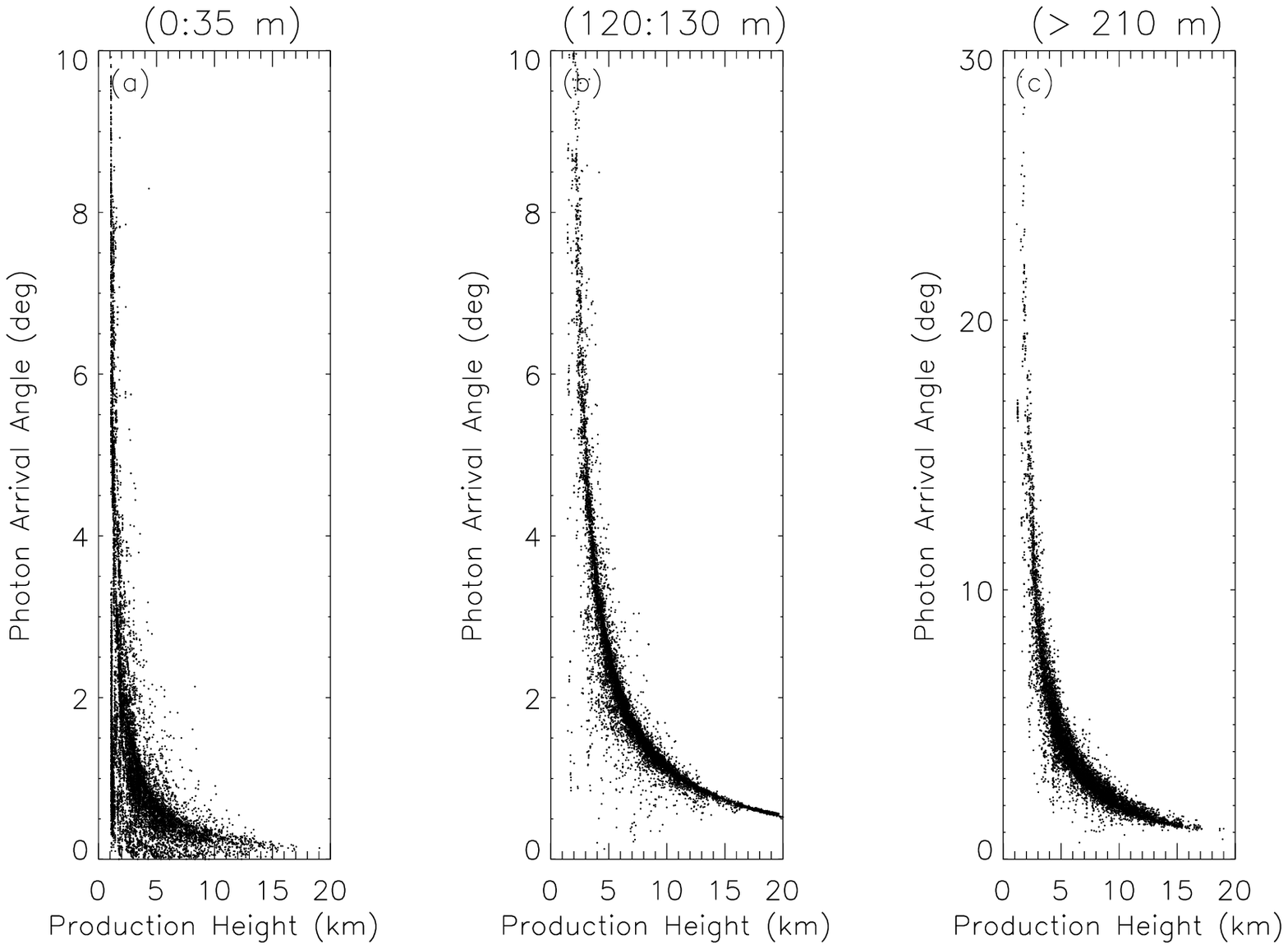,height=12cm}}
\caption{A plot of the \v Cerenkov photon arrival angle as a function of 
production height, at three different core distance ranges: pre-hump, hump 
\& post hump regions. The primary is a gamma ray of energy 550 GeV incident 
vertically on the top of the atmosphere.} 
\end{figure}

Qualitatively, the temporal radial profiles are largely independent of the
species. It can be seen from figures 2 \& 3 that the arrival time delay
increases at larger core distances giving rise to longer tails in the post
hump region.  How do we understand this? The increase in the average delay as
a function of core distance reflects the fact that moving to the outer regions
of the shower, the average energy of the electrons, the progenitors of the 
\v Cerenkov photons, decreases and therefore the deviations from a straight
line trajectory become more important. Figure 13 shows a plot of the
photon arrival time as a function of their production height, at various core
distance ranges. It can be seen that at near core distances the \v Cerenkov
photon arrival
times are almost directly proportional to their production heights and are 
produced lower down in the atmosphere.  However around the hump region,  the
light reaches the observation level first from higher altitudes and a
majority of the light can be seen to originate higher up in the atmosphere.
At larger core distances however a bulk of the photons are produced lower
down in the atmosphere consistent with the simplified picture of Hillas
[21]. A similar plot of arrival angles as shown in figure 14 shows that
the photons produced at lower heights arrive at larger incident angles
resulting mainly from multiple Coulomb scattering. Also the photons at hump
and the region beyond have large arrival angles thus leading to the
increased arrival times as seen. Hence beyond a core distance of $\sim$ 50
m, the light arrival time directly relates to their altitude of origin.

It has been shown by Hammond {\it et al.} [4] that for a vertical shower
the time taken by a photon emitted at a height {\it R} to reach a core
distance {\it r} follows a quadratic function which is approximated to a
function shown in equation 1 for a fixed height R from the observation
level. Hence the spherical shape of the \v Cerenkov light front is purely
a geometric effect, accentuated by the fact that a bulk of the \v Cerenkov
light is produced at the shower maximum. Therefore the shape of shower front
is primarily determined by the height of shower maximum and is practically
independent of species as well as altitude as confirmed by the present
results as well as by those of Protheroe {\it et al.} [22].

\subsection{Pulse shape parameters}

Both lognormal and $\Gamma -$function fits to the arrival time delay
distributions of shower secondaries have been tried by others as well.
Fitting a lognormal distribution to \v Cerenkov pulse profiles is done 
here for the first time.
Battistoni {\it et al.}[11] fitted these functions to the delay distributions of
secondary $\gamma$'s and ${\it e^\pm}$ from primary $\gamma -$rays and
protons. They conclude that lognormal distribution fits the data better
than the gamma function mainly because of the long tails at large delays
($\sim$ 200 ns). At larger delays even the lognormal function does not fit
the tail adequately and hence they use additional terms in the
Graham-Charlier expansion consisting of a series of derivatives of the
standard lognormal function.  Rodr\' iguez-Fr\' ias {\it et al.}[23] on
the other hand use $\Gamma$-function to fit the average \v Cerenkov pulses
generated by primary $\gamma -$rays and hadrons of energy $>$ 1 TeV. They
find that the function gives a good fit atleast for photon delays $<~30
~ns$, even though no mention is made of the quality of fit obtained by them.

In the present work we consider only delays $<~30~ns$. Hence we find the
quality of fit to the delay distributions to the two functions is
comparable even though lognormal function, which has a slight edge, is
used for further studies based on the fit. As shown in the Appendix B one
can derive the observable pulse shape parameters like the rise time, decay
time and the pulse width from mean pulse arrival time and the arrival time
jitter which are far more easier to measure. 

\begin{table}
\caption{Summary of the \v Cerenkov photon arrival angle distribution at the 
observation level for three different primary energies. The primary is a 
$\gamma -$ray incident vertically at the top of the atmosphere. The statistical
parameters listed below are derived from all the detected photons over all 
core distances in the range $0~-~280~m$.  The mean is computed over
50, 20 and 10 showers respectively for 100, 500 \& 1000 GeV primaries.
} 
\vskip 0.5cm
\begin{tabular}{llllll}
\hline
Energy &  Mean  & RMS  & Skewness \\
GeV    &            &            &      \\
\hline
 100   & 0.87 & 0.47 & 1.06  \\
 500   & 1.73 & 0.69 & 1.39  \\
1000   & 1.91 & 0.75 & 1.46  \\
\hline
\\
\end{tabular}
\end{table}

It has been argued in the past that the near sphericity of the \v Cerenkov
light fronts requires that the pulse shape parameters should have a
quadratic dependence on core distance [24]. Our
results for photonic as well as hadronic primaries are consistent with
this hypothesis for core distances beyond $\sim 150~m$. A steeper
dependence seen at higher observation levels could be the result of
shorter distance between the shower maximum and the observation level.
Average pulse shape parameters exhibit a minimum at a core distance of
around $150~ m$,  which is the position of the hump for $\gamma -$ray
primaries, consistent with earlier studies [21].  It is expected
since the photons arriving here are produced mainly by energetic electrons
in the cascade [20,21].  However it may be noted
that for a given primary the observed variation at short core distances
($\le $ 150 m) is comparable to the shower to shower fluctuations of these
parameters. 

It may be noted that the main reason for the asymmetric pulse shapes discussed
here is due to the large acceptance angles (unrestricted) of incident photons.
After limiting the detector opening angle to $\pm 1^\circ $ with respect to
the shower axis, it has been verified that the \v Cerenkov pulse profiles 
become narrower and more symmetric. This demonstrates that the long tails 
are contributed by photons incident at large angles. 
Cabot {\it et al.}[18] generate the \v Cerenkov pulse profile for  
vertically incident protons of energy 10 TeV which is in agreement with the
present results for lower energy protons as shown in figure 3. It can be seen
from their figures 2 \& 3 
that by applying an angular cut of $\pm 20~ mrad$ for the incident \v Cerenkov
photons, the pulse profile becomes narrower and symmetric while a bulk of the
tail at large delays vanishes.
For the same reason the average \v Cerenkov pulse profiles generated by 
Roberts {\it et al.} [19] using monte carlo technique, are comparatively 
narrower and less skewed. The main reason for this is that the opening angle of
the \v Cerenkov telescopes for which simulations are 
carried out by them is around $1.3^\circ $ FWHM.  In addition, these simulations
are carried out for higher energy {\it (50 TeV)} primaries incident at large 
zenith angles $(35^\circ~\&~65^\circ )$. More recently, the pulse shape generated
for a limited field of view of the HEGRA imaging telescope by He\ss~ 
{\it et al.}
[25] also is narrower and symmetric, once again supporting the above argument. 

Pulse shape parameters, which represent the cascade development
characteristics, are not expected to be sensitive to primary energy since
the development of showers in the atmosphere is initiated by the first
interaction point which, in turn, is determined by the interaction length 
(for a given primary species). From the present simulations we find that for 
for both types of  primaries,
the weak dependence of rise time, fall time and FWHM on the primary energy
can be parameterized as a power law in energy as shown in figure 9.
For $\gamma -$ray primaries there is a weak but definite energy dependence 
showing that each of the three pulse shape parameters increase with primary 
energy. This is mainly because the photon arrival angle distribution exhibits
larger fluctuations and becomes more skewed at higher primary energies as 
shown in table 7.
For proton primaries, on the other hand, the dependence is uncertain 
because of large errors in the fitted parameters. Considering the errors on
the slopes, our results are consistent with the expected energy independence.
However because of the logarithmic increase in the interaction cross-section 
with energy, a mild increase with primary energy is expected.  Considering
the large errors on the fitted parameters and the limited energy range
considered here, we can conclude that these fits are 
consistent with the simulation results at higher primary energies which show 
that the pulse shape parameters do increase with energy [4,23]. 

\section{Conclusions}

The radius of curvature of the \v Cerenkov shower front shows a strong
correlation with the height of shower maximum and is practically
independent of other observational parameters. \v Cerenkov light pulse
shape measurements now play an important role in the analysis of
atmospheric \v Cerenkov data and as independent measures within the
shower, complement photon density measurements. Within the range of delays
considered here lognormal distribution function represents the pulse shapes
fairly accurately at all core distances up to a maximum of $280~ m$. Among the
three pulse shape parameters considered here fall time seems to be most
sensitive to the primary species.

We would like to acknowledge the fruitful discussions with Profs. K. Sivaprasad,
B. S. Acharya and P. R. Vishwanath during the present work.

\appendix{\bf \Large{Appendix A: $\Gamma -$ function}}

 A $\Gamma -$ function distribution, {\it f(x)} is defined as [17]:
  
\begin{equation}
f(X)={{a \left[aX \right]^{b-1} exp(-aX)} \over {\Gamma(b)}}
\end{equation}

  where $X$, $a$ and $b$ are real positive numbers.

 The expectation value of {\it X} is given by:

\begin{equation}
 E(X) = {{b} \over {a}}
\end{equation}

and the variance of $X$ is given by:

\begin{equation}
V(X) = {{b} \over {a^2}}
\end{equation}

Maximum functional value is obtained by solving the equation: 

\begin{equation}
{{df} \over {dX}} = 0 
\end{equation}

The solution for which is 

\begin{equation}
X_{max} = {{b-1} \over {a}}
\end{equation}

$a$ and $b$ in the $\Gamma -$  function can be expressed in terms of peak 
position $X_{max}$ and the mean value $<X>$ using the equations given above. 
Hence $\Gamma -$ function could be defined in terms of the newly chosen
variables as:

\begin{equation}
f(X) = C \times X^{\left({X_max} \over {<X> - X_{max}}\right)} exp{\left(-{{X} \over {<X>-X_{max}}} \right)}
\end{equation}

where $$ C={{a^b} \over {\Gamma(b)}}$$

However, in order to use this expression it is necessary to know the
position of the pulse peak. Hence it is necessary to suitably bin the
arrival times at the detector and generate a pulse profile. The estimate of
the pulse peak position can therefore be bin dependent. We have tried to
avoid this dependence on binning, by obtaining the initial estimates of pulse 
parameters after
expressing $a$ and $b$ in terms of expectation value [$E(X)$] and variance
[$V(X)$]. These
are arithmetic mean and variance of the data, derived using equations given 
above.  Hence

\begin{equation}
a={{E(X)} \over {V(X)}}
\end{equation}

and

\begin{equation}
b=a \times E(X)
\end{equation}

These values of $a$ and $b$ are used as initial estimates while fitting pulse
profile to a Gamma distribution function. However one cannot easily derive the 
pulse shape parameters like rise \& fall times and full width of the pulse,
using $a$ and $b$.

\appendix{\bf \Large{Appendix B: Lognormal distribution}}

 A positive variate $x$ $(0<x<\infty)$ follows a lognormal distribution
when its logarithm $y=logx$ is normally distributed with mean $\mu$ and
variance $\sigma^2$.  Distribution function is given by [26]: 

\begin{equation}
P(x) = {{1} \over {x \sigma \sqrt{2 \pi}}} exp{\left[-{{1} \over {2 \sigma^2}} \left(log x -\mu \right)^2 \right]}
\end{equation}

The $j$th moment of the distribution about the origin is given by:

\begin{equation}
\lambda'_j = exp \left(j \mu + {{1} \over {2}} j^2 \sigma^2 \right)
\end{equation}

Hence the mean $\alpha$ and variance $\beta^2$ are given by

\begin{equation}
\alpha = exp \left(\mu + {{1} \over {2}} \sigma^2 \right) 
\end{equation}

and

\begin{equation}
\beta^2 = exp(2 \mu + \sigma^2) \left(exp(\sigma^2)-1\right)
        = \alpha^2 \eta^2
\end{equation}

where  $\eta^2 = exp(\sigma^2)-1 $

For a given data set arithmetic mean $\alpha$ and variance $\beta^2$ can
be estimated and using the equations given above. $\mu$ and $\sigma^2$
can be calculated after inverting the equations 3 \& 4 as:

\begin{equation}
\mu = {log {\alpha}} - {{1} \over {2}} {\sigma^2}
\end{equation}
 
\begin{equation}
\sigma^2 = log \left( 1 + {{\beta^2} \over {\alpha^2}} \right) 
\end{equation}

and then the corresponding lognormal distribution can be
generated. The values of $\mu$ and $\sigma^2$ so obtained are used as
initial estimate while fitting lognormal distribution to pulse profile.

Median and mode of the distribution are given by $exp(\mu)$ and 
$exp(\mu -\sigma^2)$, respectively.

Value of the distribution function at the mode is given by

\begin{equation}
P_{max} = {{1} \over {\sigma \sqrt{2 \pi}}} exp \left[{{\sigma^2} \over {2}} -\mu \right]
\end{equation}

FWHM of the distribution, i.e, the width where value of the distribution 
function is $P_{max}/2$, can be obtained solving equation 1, after substituting 
for $P_{max}$ from equation 7. It is given by,

\begin{equation}
w = exp \left(\mu - \sigma^2 \right)  \left[ exp \left( \sqrt{-2 \sigma^2 log(0.5)} \right) - exp \left( -\sqrt{-2 \sigma^2 log(0.5)} \right) \right]
\end{equation}

Rise time and decay time can be calculated using a similar procedure.
They are given by following expressions:

\begin{equation}
\tau_r = exp \left(\mu - \sigma^2 \right) \left[ exp \left( -\sqrt{-2 \sigma^2 log(0.9)} \right) - exp \left( -\sqrt{-2 \sigma^2 log(0.1)} \right) \right]
\end{equation}

\begin{equation}
\tau_d = exp  \left(\mu - \sigma^2 \right) \left[ exp \left( \sqrt{-2 \sigma^2 log(0.1)} \right) - exp \left( \sqrt{-2 \sigma^2 log(0.9)} \right) \right]
\end{equation}

\end{document}